\documentclass[12pt,a4paper]{article}
\usepackage[T2A]{fontenc}
\usepackage[cp1251]{inputenc}
\usepackage[english]{babel}
\usepackage{amssymb}
\usepackage{amsmath}
\usepackage{graphicx}
\usepackage[small,sc,center]{titlesec}
\usepackage{indentfirst}
\usepackage[labelsep=period]{caption}
\usepackage{caption}
\newcommand{\msun}{\,$M_{\odot}$}

\newcommand{\lsun}{\,$L_{\odot}$}
\newcommand{\ergs}{\,erg\,s$^{-1}$}
\newcommand{\gcmq}{\,g\,cm$^{-3}$}

\newcommand{\gcmsq}{\,g\,cm$^{-2}$}
\newcommand{\kms}{\,km\,s$^{-1}$}

\newcommand{\cmq}{\,cm$^{-3}$}

\newcommand{\arii}{Ar\,{\sc ii}}
\newcommand{\arvi}{Ar\,{\sc vi}}
\newcommand{\mcm}{\,$\mu$m}
\newcommand{\mcg}{\,$\mu$G}
\begin{document}
	
\begin{center}
\textbf{\large Puzzle of [\arii]\,7\,\mcm\ line broad component of SN~1987A}
\vskip 5mm
\copyright\quad
2024 \quad N. N. Chugai\footnote{email: nchugai@inasan.ru} and V. P. Utrobin$^{2,1}$\\
\textit{$^1$ Institute of astronomy, Russian Academy of Sciences, Moscow} \\
\textit{$^2$SRC ``Kurchatov institute'', Moscow} \\
Submitted  05.10.2023 
\end{center}


\noindent
{\em PACS codes:\/} 

\clearpage
 
 \begin{abstract} 
 	
We explore the origin of the broad component of the [\arii] 7\,\mcm\ line emission 
   related to the ejecta excitation by the neutron star of SN~1987A.
We argue that the line broad wings are emitted at the tmperature of $\sim300$\,K. 
 The flux excess in the red wing of [\arii] line is reproduced by the line photons scattering off the optically thin uniform dust component with the grain size of 1 - 2\mcm\ and the total mass of $\mbox{(several)}\times10^{-3}$\msun.
 The dusty opaque clumps containing almost all the dust of SN~1987A have a low occultation optical depth and line photon scattering on dusty clumps do not contribute noticeably in the red wing. 
The additional heating might be related to ionization losses of    
 relativistic protons.
\end{abstract}

\section{Introduction}

The idea that emission lines of high ionization species can provide evidence for pulsar excitation in SN~1987A even though the  
   pulsar is not directly observed (Chevalier \& Fransson 1992)    
   motivated observation of SN~1987A by JWST on 16 July 2022  (Fransson et al. 2024).
The obtained infrared spectra show narrow lines of argon and  
  sulfur ions  that originate from gas illuminated by the ionizing radiation of a cooling neutron star (CNS) or a pulsar wind nebula  (PWN) (Fransson et al. 2024). 

Of a special interest are [\arii] 6.98527\,\mcm\ and [\arvi] 4.5\,\mcm\ lines. 
The latter is a signature of the hard ionizing radiation related to the NS (Fransson et al. 2024),  while the  former line with a high signal-to-noise ratio permits ones to explore the structure of  the line-emitting  gas.
It was emphasised that the narrow core of [\arii] 7\,\mcm\ favours the dominant role of the gas  
  ionization by the CNS in contrast to the scenario of the PWN (Fransson et al. 2024). 
   
The indirect, yet convincing, detection of NS based on spectral observation poses a question, whether 
  we understand the relevant physics of the [\arii] 7\,\mcm\ line formation. 
At least two principal issues require further elucidation.
First,  the model of the ionization and thermal state of gas excited by the CNS 
  (Fransson et al. 2024) is able to account only for the narrow component (FWHM $\sim 120$\kms) of the [\arii] 7\,\mcm\ line. 
This leaves unexplained the origin of the broad (FWHM $\sim 360$\kms) component.
 Second, the [\arii] 7\,\mcm\ line shows the spectacular flux excess  in the red wing,
  which could be related to the line photons scattering off dust (Fransson et al. 2024), although the conjecture requires confirmation by a numerical modelling.
Noteworthy, effects of photon scattering off the dust can be hampered by the fact that the dust 
  resides in opaque clouds (Dwek \& Arendt 2015).  

The present communication explores both raised issues.
The study is based on a simple spherical model with the line emissivity distribution   
 implied by the line profile.
While the model is rather simple, it permits us to infer interestong conclusions on the  dust distribution and thermal state of the 
 line-emitting gas responsible for the broad component of the [\arii] 7\,\mcm\ line. 
 
\begin{table}
	\vspace{6mm}
	\centering
	{{\bf Table 1.} Model parameters. }\\
	\bigskip	 
	\begin{tabular}{p{1.5cm}|c|c|c} 
		\hline	
		Model &   $\tau_d$   & $\omega$ &  $g$  \\
		\hline	
		A     &   0     &  0       &  0 \\
		B    &   0.9    &  0.5     &  0 \\
		C     &    0.9   &   0.5    &  0.3 \\
		D    &    0.9\,$^a$   &   0.5\,/0.11\,$^b$ & 0 \\
		\hline
	\end{tabular}
	\medskip
	
	\begin{tabular}{l}
		$^a$ $\tau_{oc}\qquad ^b$ $\omega$/cloud albedo\\	
	\end{tabular}
\end{table}

  \section{Line profile modelling}
  
  \subsection{Model overview}
  
The line profile of [\arii] line is calculated assuming the spherical line-emitting zone around the NS with the 
 emissivity $\epsilon (r)$ and a  homologous expansion $v = r/t$, where $r$ is the distance from the NS and $t = 35$\,yr is the supernova age at the JWST observation. 
Noteworthy, the homologous expansion with the center at the NS  position is a natural consequence of the overall homologous expansion of ejecta; we neglect velocity field perturbations by the PWN (if any).  
Below we predominantly mean the gas velocities with respect to the NS, except for  
  cases when we consider a velocity with respect to the overall ejecta expansion center.
 
Scattering and absorption of line photons by the dust is calculated based on the Monte Carlo technique assuming 
  uniform dust distribution in the sphere of the boundary velocity $v_d = 800$\kms\ with the center at the NS. 
Note, $v_d$ value is close to the velocity of dusty ejecta 910\kms\ used for the IR spectra  
  modelling (Dwek \& Arendt 2015).  
There might be some displacement of the \arii\ line-emitting gas  w.r.t. the dust distribution   
  in the ejecta, but we neglect the displacement that could introduce only insignificant correction in the calculated profile. 
  
The relevant dust properties include the total (i.e., extinction) optical depth of the dusty sphere $\tau_d$, the   single scattering albedo $\omega$, and the average cosine $\mu$ of the angle $\theta$ between an incident and scattered photon $g = \langle \mu\rangle$. 
The  adopted phase function proposed by Draine (2003) is 
\begin{equation}
\Phi(\mu) = C\frac{1 + \mu^2}{(1 + g^2 - 2g\mu)^{3/2}}\quad,
\end{equation}
 where $C$ is a normalizing factor.  
For small grains, viz., $x= 2\pi a/\lambda \ll 1$, parameter $g \rightarrow 0$ and this expression   
  recovers the Rayleigh phase function. 
  
The {\em Herschel}~infrared spectra at the age of 25 yr (Wessel et al. 2015) suggest the micron-sized grains with the radii up to 3\mcm. 
This implies a significant asymmetry parameter ($g > 0$)  for the 7\mcm--photons.
Below several relevant options are presented to demonstrate effects of major parameters 
 on the line  profile.
 
We explore both cases of the dust distribution ---  uniform and clumpy.
The dust in optically thick clouds for SN~1987A has been invoked to reconcile 
  the blueshift of the [O\,I]\,6300, 6364\AA\ lines that requires a moderate dust optical depth ($\tau_d \lesssim 1$) with the infrared dust emission that suggests optically thick 
    dusty zone $\tau_d \gg 1$ (Lucy et al. 1991).
The large optical depth of dusty clouds ($\tau_c \gg 1$) is the case at the large age of SN~1987A (25 yr) 
as well (Dwek \& Arendt 2015; Wesson et al. 2015) with the total mass of dust of    
   0.8\msun\  (Wesson et al. 2015). 
   
If the dust resides only in opaque clouds, the appropriate parameter is 
the "occultation optical depth" ($\tau_{oc}$) (Lucy et al. 1991), which is equivalent of the 
   average number of clouds crossed by the radius of dusty zone. 
The photon scattering off dusty opaque cloud is a tricky issue, since the shape of a cloud surface, as a rule, is rather complicated.
We simplify the problem assuming a diffuse Lambertian reflection of light from a smooth surface 
  assuming a normal photon incidence ($\cos{\theta_0} = \mu_0 = 1$).
 In the approximation of a large cloud optical depth $\tau_c \gg 1$ and spherical phase function 
   the flat albedo is $A = 1 - \phi(\mu_0,\omega)\sqrt{1 - \omega}$  (Sobolev 1985).
The function $\phi(\mu_0,\omega)$ for $\omega = 0.5$ (astrosilicate 1\mcm\ grain at $\lambda = 7$\mcm\ (Draine 2011)) and $\mu_0 = 1$ is equal 1.26 (Sobolev 1985), so 
the flat albedo  is  $A = 0.11$ that is used below.

\begin{figure}
	\centering
	\includegraphics[width=0.9\columnwidth]{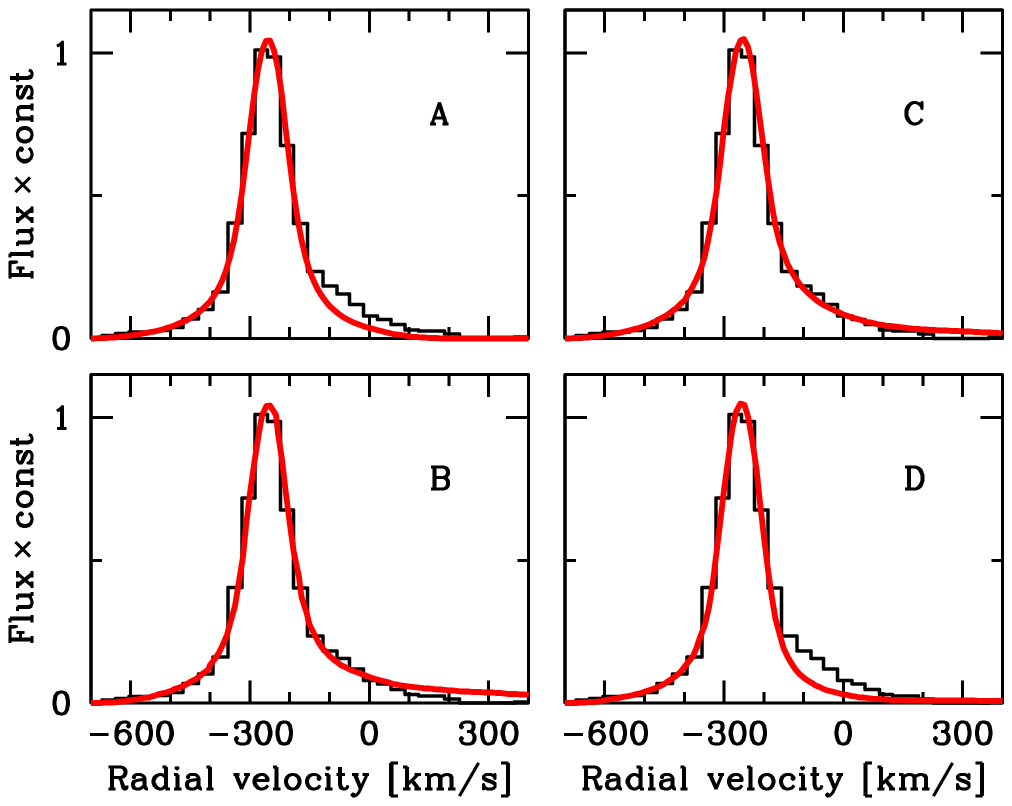}
	\caption{
		\small{
			Model line profile (red line) for  cases A, B, C, and D (Table 1) overploted on the observational 
			[\arii]\,7\,\mcm\ line profile. Model C is preferred compared to model B
			with the stronger far red wing; 		
			model D with all the dust in clouds fails to describe the red wing.}
	}
	\label{fig:prof}
\end{figure}

  \subsection{Optically thin dust component}
  
Model profiles for different parameters (Table) alongside with the observed 
 line are presented in Figure 1.
Calculated profiles are blueshifted by -260\kms\ to allow for the neutron star 
 radial velocity  (Fransson et al. 2024).
All the models use the same normalized emissivity $\epsilon(v)$  with the 
  velocity $v = r/t$ with respect to the NS (Figure 2). 
  
Model A with zero dust optical depth shows the significance of the photon scattering 
  off the dust in the formation of the red wing.
The enhanced red wing appeares in models B and C with the same dust optical depth but   
 different asymmetry parameter.
Model C with $g = 0.3$ is preffered compared to  model B with $g = 0$. 
In the latter the far red wing ($v_r \gtrsim +50$\kms) is unacceptably strong compared to 
 the observed line profile. 
Model D with the photon scattering off opaque dusty clouds with the occultation optical depth  $\tau_{oc} = 0.9$ is unable to account for the red wing at all and should be rejected.
We do not show models with different value of the radius (velocity) of the 
 dusty sphere, yet the value $v_d = 800$\kms\ is found to be the optimal one with the 
 uncertainty of +100/-200\kms.
 
The major conclusion of the profile modelling is that all the 
 dust should not be locked in opaque clouds. 
There must be an intercloud smoothly distributed dust with the optical depth of $\approx 0.9$  
   at the velocity scale of $r/t \approx 800$\kms.
Furthermore, the ensemble of dusty clouds should have the low occultation 
 optical depth in order not to affect the line profile, i.e.,  $\tau_{oc} = (3/4)(R_d/r_c)f < 1$, where $f$ is the filling factor of the cloudy component, $R_d = v_dt \sim10^{17}$\,cm is the radius of a cloudy zone, and $r_c$ is a typical cloud radius        .
To illustrate the point, the occultation optical depth of 0.5 for $R_d/r_c = 50$ requires the filling factor $f \sim 0.01$.
 
The total mass of the homogeneous optically thin dust is 
\begin{equation}
M_d = 2\times10^{-3}\zeta\left(\frac{\tau_d}{0.9}\right)\left(\frac{v_d}{800\mbox{\kms}}\right)^2\left(
\frac{a}{1\mbox{\mcm}}\right)\,M_{\odot}\,,
\end{equation}
where $\zeta \approx 1$\gcmq\ is the density of the grain material.
Interestingly, while the micron-sized grains are implied by the  
 infrared spectral energy distribution (Wesson et al. 2015), the large grains 
  are required as well by the asymmetry parameter $g = 0.3$  of optimal model C to describe 
  red wing of \arii\ line. 

Indeed, according to Mie theory, spherical grains with the parameter $x = 2\pi a/\lambda$  
 in the range $1 < x < 2$ have the asymmetry parameter $0.2 < g < 0.6$ (Ehlers \& Moosm\"{u}ller 2023).
The interval $1 < x < 2$ for $\lambda = 7$\mcm\ corresponds to the  grain size 
   $1.1 < a < 2.2$\,\mcm.
The grain size showing asymmetry $g = 0.3$ thus lies in the range of 1 - 2 \mcm. 
Adopting $a = 1.5$\mcm, one obtains from Equation (2) the total mass of the uniform dust component of $\sim 0.003$\msun.
With the absorption efficiency of astrosilicates $Q_{abs} = 0.014(a/1\mbox{\mcm})(100\mbox{\mcm}/\lambda)^2$ (Draine 2011) and the fiducial grain temperature of 20\,K (Wesson et al. 2015) the luminosity of the uniform dust component 
 is $4$\lsun, which is $\sim 2$\% of the core dust luminosity (Wesson et al. 2015). 
 
To conclude, the optically thin uniform dust component responsible for the red 
  wing of [\arii]\,7\,\mcm\ line is made of 1 - 2\mcm\ grains with the total mass of $\sim \mbox{(several)}\times10^{-3}$\msun.
 One cannot rule out that the uniform component of dust is composed by rarefied halo of dusty clumps.

 \subsection{Broad component and electron temperature}
 
 The [\arii]~7\,\mcm\ emission is considered as composed of two Gaussian components:
  narrow (FWHM = 120\kms) with the luminosity  $L_n = 1.4\times10^{32}$\ergs\ and broad (FWHM = 360\kms)
   with the luminosity $L_b = 10^{32}$\ergs\  (Fransson et al. 2024).
  Half of the latter in line with our modelling we attribute to the narrow component scattered off the dust, 
  so the original broad component comprises  $\approx 5\times10^{31}$\ergs.
  
The broad component is emitted by the gas with velocities of 60 - 350\kms\ (we ignore tangential velocity 
 that is known with a large error), i.e., at distances $(6 - 40)\times10^{15}$\,cm from the NS. 
The CNS and PWN ionization models consider the line-emitting gas in the velocity range of
   $\lesssim 70$\kms\ (Fransson et al. 2024).
In both models the ionization fraction and the electron temperature decrease with  
  distance from the NS. 
 Extrapolating these models in the velocity region $\gtrsim 70$\kms\ 
  we adopt the ionization fraction $x_e = 0.03$, \arii\ fraction of 0.03 and $T_e = 100$\,K.
   
 The emissivity $\epsilon(v)$  of [\arii]\,7\mcm\ line ($J_1 = 3/2$, $J_2 = 1/2$, $A_{21} = 5.3\times10^{-2}$\,c$^{-1}$, $T_{ex} = E_{12}/k =2060$\,K),  
 is determined by the electron collisions. 
Excitation by colisions with the neutral oxygen is plausible, but the rate is unknown.
Based on the \arii\ excitation by neutral hydrogen with the rate of four orders smaller than 
 for elctron collisions (Yan \& Babb 2024) we neglect collisions with the neutral oxygen.

Following (Fransson et al. 2024), we adopt Ar abundance of $2.7\times10^{-2}$ by number  
for the  O-Si-S-Ar-Ca  mixture and the average 
ion (and neutrals) density $n_i = 2.7\times10^3$\cmq\ for 3\msun\ ejecta core in the  
 ejecta velocity range $v< 2200$\kms.  
For the collisional strength  $\Omega_{21} = (2J_1+1)(2J_2+1)$ (Gould 1963) 
 the rate coefficient of the collisional transition is $q_{21} = 3.4\times10^{-6}(T_e/100K)^{-0.5}$\,cм$^3$\,с$^{-1}$.  
The electron number density ($\lesssim 10^2$\cmq) is significantly lower than the 
critical density $A_{21}/q_{21} \approx 10^4$\cmq, so the emissivity is
\begin{equation}
	\epsilon = E_{12}(g_2/g_1)q_{21}n_en(\mbox{\arii})\exp{(-T_{ex}/T_e)}\,.
\end{equation}
  where $n(\mbox{\arii})$ is the number density of \arii.
For adopted density, ionization, abundance and temperature of 100\,K 
the luminosity of the broad component of [\arii]\,7\mcm\
in the line-emitting zone ($< 350$\kms) turns out to be 
 $L_b \approx 2.4\times10^{25}$\ergs, six dex lower 
  compared to the observational value.
This is the outcome of the extremely low exponential factor, which indicates that 
 electron temperature of the line-emitting gas with velocities $> 60$\kms\ 
  exceeds adopted temperature of 100\,K. 
For the same density and ionization fraction, the "average" electron temperature that is 
 needed to reproduce the luminosity of the broad component is $T_e \approx 390$\,K.
The line-emitting gas responsible for the broad component thus should be significantly 
 warmer compared to that in the extrapolated CNS and PWN models. 

The exponential factor essentially determines the $\epsilon(v)$ behavior. 
This provides us with an interesting possibility to estimate independently the temperature   
  in the high velocity region ($v \sim 100$\kms) using the line profile modelling.
In the presented line profile models we adopt the power law for the emissivity behavior 
  $\epsilon \propto v^{-\beta}$ with $\beta = 3.5$  in the region $v > 60$\kms.
For the constant ionization fraction and  the power law $T_e \propto v^{-\gamma}$,  
  the logarithmic differentiaton of  
   the  expression $\epsilon \propto T_e^{-1/2}\exp{(-T_{ex}/T_e)}$ results in  
   the simple formula for the electron temperature 
\begin{equation} 
	T_e = T_{ex}\left(\frac{\beta}{\gamma} + 0.5\right)^{-1}\,.
\label{eq:teme}	
\end{equation}
This expression is relevant for the velocities $\sim 100$\kms.
The $\gamma$ value can be estimated numerically by the fitting $\epsilon(v)$ recovered  
  earlier from the line profile.
  
The fittng of $\epsilon(v)$ behavior using relations  $\epsilon \propto T_e^{-1/2}\exp{(-T_{ex}/T_e)}$, $T_e = T_0(v/v_0)^{-\gamma}$ (where $v_0 = 100$\kms), and 
 $\epsilon \propto v^{-\beta}$ with $\beta = 3.5$  is shown in Figure 2. The visual best fit is found for $\gamma = 0.5$ and $T_0 = 350$\,K. The latter is close to the 
  estimated temperature of 390\,K based on the broad component luminosity. 
On the other hand, the Equation (\ref{eq:teme}) for $\gamma = 0.5$ suggests 
 a bit lower value $T_e = 275$\,K.
 
We conclude that the electron temperature in the high velocity gas $v \approx 100$\kms\ 
   lies in the range of 300 -- 400\,K and slowly decreases with the velocity as
    $T_e \propto v^{-0.5}$.
If the ionization fraction decreases with  the expansion velocity, one expects  
 that $\gamma < 0.5$  in the line-emitting zone responsible for the broad component.

\begin{figure}
	\centering
	\includegraphics[width=0.9\columnwidth]{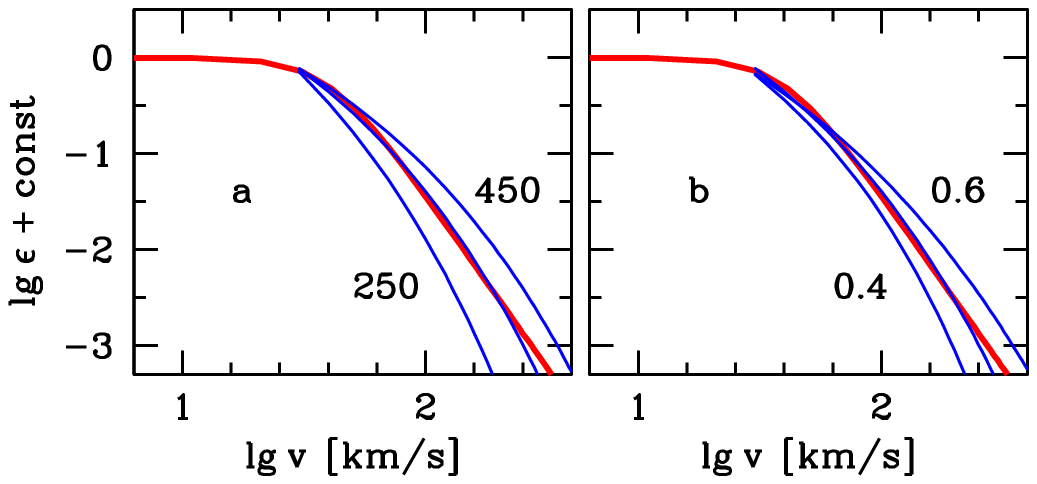}
	\caption{
		\small{Emissivity distribution in the model of [\arii] line ({\em red}) with the high velocity fit 
			by temperature distribution  $(T_e/T_0) = (v/v_0)^{-\gamma}$. Panel {\bf a} shows the case of different $T_0$ (250, 350, 450\,K) with fixed $\gamma = 0.5$
			; panel  {\bf b} shows the case of different $\gamma$ (0.4, 0.5, 0.6) 
			with fixed . Best fit corresponds to $T_0 = 350$\,K and $\gamma = 0.5$.				}
	}
	\label{fig:emis}
\end{figure}

\section{Discussion and conclusions}
	
The aim  of this paper has been to explore the origin of the broad component of 
 the [\arii]~7\,\mcm\ line emission from SN~1987A core at 35 yr and to account for 
  the flux excess in the red wing of this line.
We conclude that the enhanced red wing is the outcome of the line photons scattering off the 
  optically thin quasi-uniform dust component with the grain size of 
   1 - 2\mcm\ and the total mass of $\sim \mbox{(several)}\times10^{-3}$\msun. 
We rule out the dominant contribution of the line photon scattering off opaque dusty clouds.  
  
In the proposed scenario the broad line component 
 originates from the gas with expansion velocities 
 $60 < v < 350$\kms\ (w.r.t. the NS) and the electron temperature in the range of 300-400\,K.   
This value is significantly higher than the temperature $\lesssim 100$\,K predicted by 
the  CNS and PWN models (Fransson et al. 2024) and extrapolated to velocities $> 70$\kms.
  
A question may arise whether the cold dust $T_d \approx 20$\,K (Wesson et al. 2015) can coexist with the  warm  $T_e \approx 350$\,K  environment?
We estimate that the grain total heating rate by elecrons, ions,  and neutrals
 is $\approx 30$ times lower compared to the radiative cooling of the grain.
This remove doubts concerning the coexistence of the cold dust and the warm gas.  

A challenging issue is the origin of an extra heating source for 
 the broad component of [\arii] line. 
Positrons from $^{44}$Ti decay are trapped in  Fe-clumps as formerly suggested 
 for 8 year-old SN~1987A (Chugai et al. 1997). 
Positrons should be trapeed at present as well, otherwise they would produce unobserved   
   broad line wings up to $\approx 1000$\kms. 
The heating apparently should be bounded by a close zone around the NS.
 
In the absence of an alternative, we conjecture that the additional heating could be maintained by ionization losses of relativistic protons escaping from the PWN. 
Currently the lepton-to-proton ratio in the PWN is an open issue, since 
   the signature of relativistic protons is hard to observe.
The detection of PeV photons from  the Crab nebular by the LHAASO facility indicates that relativitic protons are there and they 
carry 10--50\% of the NS spindown luminosity with most of relativistic protons escaping the PWN (Liu \& Wang 2021).

This heating mechanism suggests that the diffusion time of protons in the sphere of 
a radius $R\sim 2\times10^{16}$\,cm should be an order of the supernova age $t = 35$\,yr.
This means that the diffusion coefficient should be comparable to the supernova age $t_{dif} \sim R^2/6D \sim 35$ yr.
A diffusion time of the relativistic particle ($\gamma \gg 1$) in the 
  magnetic field is similar for  electron and proton of the same energy. 
However for the energy $\gtrsim 100$ MeV the electron ionization losses are
 negligibly small compared to radiation losses, so the heating by protons  
 turns out more efficient. 
In the case of the Bohm diffusion, $D = r_gc/3$, were $r_g = E/eB$ is a giroradius, the particle energy $E$ and magnetic field are constrained as 
 $B \sim 5(E/10\mbox{\,GeV})(R/2\times10^{16}\mbox{\,cm})^{-2}$\,\mcg; for $B \approx 5$\mcg, a proton energy should be $\sim 10$\,GeV.
Note, $B \approx 5$\mcg\ is also needed to inhibit the escape of $\sim 1$ MeV positrons of the 
 $^{44}$Ti decay from Fe-clumps with a size of $\sim 10^{-2}R \sim 2\times10^{14}$\,cm.
The PWN magnetic field is likely stronger and  
  has  a complicated topology that could hamper or, on the contrary, facilitate the escape of relativistic protons. 
Yet with the PWN radius (Fransson et al.2024) $\sim 10$ times smaller  than that of the line-emitting zone for 
   the broad component, the diffusion time in the PWN is comparable to that of 
   the line-emitting zone even for 100-fold  stronger field of the PWN.

The range of 10\,GeV protons due to ionization losses in the O-Si mixture  is   
   $R_{ion} \approx 5\times10^3$\gcmsq\ (NIST data).
For the density $\rho \approx 10^{-19}$\gcmq\ in the central zone of SN~1987A  
  the time of the proton energy loss is $t_{ion} = R_{ion}/(c\rho) \sim 1.6\times10^{12}$\,s.
The power of the ionization losses required to maintain additional heating of the \arii\ 
  line-emitting zone is $L_{ion}\sim 10^{32}$\ergs. 
This suggests the total energy of relativistic protons $W_p\sim L_{ion}t_{ion} \sim 1.6\times10^{44}$ erg and the number of 10\,GeV protons $N_p \sim 10^{46}$. 
The energy loss of relativistic protons is determined by $pdV$ work
 $W_p/t \sim 1.4\times10^{35}$\ergs. 
This is significantly less than the estimated upper limit of the 
 neutron star rotational energy loss  $\lesssim 3\times10^{36}$\ergs\ (Fransson et al. 2024).

Proton-nuclues collisions result in the $\pi^0$ production with the subsequent decay into 
  two gamma-quanta with the total rest-frame energy $m_{\pi}c^2 = 135$\,MeV.
A detection of gamma-quanta from $\pi^0$, or the flux upper limit, could verify the proposed 
 heating mechanism.
Given the number density  of O-Si mixture $3\times10^3$\cmq\ with the average atomic weight $A=20$, the nucleus radius $R = 1.25A^{1/3}$\,Fm and the cross section $\sigma_n =\pi R^2 = 385$\,mb,  the free flight time for nuclear collisions 
  is $t_n \sim 3\times10^{11}$\,s.
Experiments on the proton collision with nuclei of $^4$He and $^{12}$C 
(Yang et al. 2018) extrapolated to 
 nucleus with $A = 22$ show that the cross section for the $\pi^0$ production at the energy of 10 Gev 
 is $\sigma_{\pi^0}\approx 360$\,mb, close to the geometric cross section. 
 We thus conclude that each $pA$-collision for $A = 22$ produce on average one neutral pion. 
 The total luminosity of gamma-quanta (neglecting the momentum of produced $\pi^0$) is then 
  $L_{\gamma} = m_{\pi}c^2N_p/t_n \sim 10^{31}$\ergs.

In the proposed scenario gamma-ray flux at about 100 MeV given the distance to LMC of 
  50 kpc should be $L_{\gamma}/(4\pi D^2) \sim 3.5\times10^{-17}$\,erg\,cm$^{-2}$\,s$^{-1}$.
This value is by the 5 dex lower than the detection limit of Fermi LAT at 100 MeV with the   
  exposition of 4 yr ({\em https://fermi.gsfc.nasa.gov}).
The low expected flux of gamma-quanta from $\pi^0$ decay therefore leaves open the verification 
 issue for the proposed mechanism of additional heating of the gas responsible for the 
  broad component of [\arii]\,7\mcm\ line. 
We therefore consider the proposed mechanism as an interesting possibility that requires 
confirmation.

\clearpage

\section{References}

\noindent
Chevalier R., Fransson C., Astrophys. J. {\bf 395}, 540 (1992)\\
\medskip 
Chugai N., Chevalier R., Kirshner R.,  Challis P., Astrophys. J. {\bf 483}, 925 (1997)\\
\medskip 
Draine B. T. {\em Physics of the Interstellar and Intergalactic Medium.} Princeton: Princeton University Press, 2011\\
\medskip 
Draine B. T., Astrophys. J. {\bf 598}, 1017 (2003)\\
\medskip 
Dwek E., Arendt R. G., Astrophys. J. {\bf 810}, 75 (2015)\\
\medskip
Ehler K., Moosm\"{u}ller H., Aerosol Science and Technology {\bf 57}, 425 (2023)\\
\medskip 
Fransson C., Barlow  M. J., Kavanagh P. J. et al., Science {\bf 383}, 898 (2024)\\
\medskip
Gould R. J., Astrophys. J. {\bf 138}, 1308 (1963)\\
\medskip
Liu R.-Y., Wang X.-Y. Astrophys. J. {\bf 922}, 221 (2021)\\
\medskip
Lucy L. B., Danziger I. J., Gouiffes C., Bouchet P. 
{\em Supernovae. The Tenth Santa Cruz Workshop in Astronomy and Astrophysics.}
Ed. S.E. Woosley. New York: Springer-Verlag, 1991\\
\medskip 
Sobolev V. V. {\em Light scattering in planetary atmospheres.} 
 Oxford and New York: Pergamon Press, 1975\\
\medskip 
Wesson R., Barlow M. J., Matsuura M., Ercolano B., Mon. Not. Royal Astron. Soc. {\bf 446},
 2089 (2015)\\
\medskip
Yan P.-G., Babb J. F., Astrophys. J. {\bf 961}, 43 (2024)\\
Yang R., Kafexhiu E., Aharonian F., Astron. Astrophys. 
{\bf 615}, A108 (2018)\\

\end{document}